%% file: aaskaii_template.tex
\documentclass[a4paper,11pt]{article}
\usepackage{aaskaiid}
\usepackage{orcidlink}
\usepackage{sidecap}
\title{Gamma-ray Bursts in the Radio Sky: the Role of the SKA-VLBI}
\ShortTitle{GRBs in the Radio Sky: the Role of the SKA-VLBI}

\author[1]{Stefano Giarratana\orcidlink{0000-0002-2815-7291}}
\ShortName{Giarratana, S. et al.} 
\author[1,2]{Om S. Salafia\orcidlink{0000-0003-4924-7322}}
\author[3,4]{Xue-Feng Wu\orcidlink{0000-0002-6299-1263}}
\author[3]{Jin-Jun Geng\orcidlink{0000-0001-9648-7295}}
\author[5,6]{Tao An\orcidlink{0000-0003-4341-0029}}
\author[7,8]{Benito Marcote\orcidlink{0000-0001-9814-2354}}
\author[9]{Marcello Giroletti\orcidlink{0000-0002-8657-8852}}
\author[1,2]{Giancarlo Ghirlanda\orcidlink{0000-0001-5876-9259}}
\author[9]{Tiziana Venturi\orcidlink{0000-0002-8476-6307}}

\affiliation[1]{INAF Osservatorio Astronomico di Brera, Via E. Bianchi 46, I-23807 Merate, Italy}
\emailAdd{stefano.giarratana@inaf.it}
\affiliation[2]{INFN -- sezione di Milano-Bicocca, Piazza della Scienza 3, I-20126 Milano, Italy}
\affiliation[3]{Purple Mountain Observatory, Chinese Academy of Sciences, Nanjing 210023, People’s Republic of China}
\affiliation[4]{School of Astronomy and Space Sciences, University of Science and Technology of China, Hefei 230026, People’s Republic of China}
\affiliation[5]{Department of Astronomy, University of Science and Technology of China, Hefei, Anhui 230026, China}
\affiliation[6]{Shanghai Astronomical Observatory, Chinese Academy of Sciences, Nandan Road 80, Shanghai 200030, People’s Republic of China}
\affiliation[7]{Joint Institute for VLBI ERIC, Oude Hoogeveensedijk 4, 7991 PD Dwingeloo, The Netherlands}
\affiliation[8]{ASTRON, Netherlands Institute for Radio Astronomy, Oude Hoogeveensedijk 4, 7991 PD Dwingeloo, The Netherlands}
\affiliation[9]{INAF Istituto di Radioastronomia, via Gobetti 101, 40129 Bologna, Italy}

\abstract{Radio observations of $\gamma$-ray bursts (GRBs) employing the very long baseline interferometry (VLBI) technique provide us with fundamental information on the dynamics and the geometry of the GRB blast wave. With its high angular resolution ($\sim$milli-arcsecond), VLBI allows us to measure the apparent superluminal expansion, to characterise the structure of the jet and to constrain the viewing angle and jet opening angle. While this information is crucial 
to understand these transient events, such studies have been possible only for three GRBs to date, 
owing to both the poor sensitivity of current radio facilities and the paucity of close and bright GRBs. In this chapter, we estimate the impact that the SKA-Mid will have on these studies, when included in a VLBI network. We performed a series of dedicated simulations of VLBI observations of GRBs, considering five VLBI networks and the SKA-Mid, both in its AA* and AA4 configurations. 
We show that 
including the SKA-Mid in a global-VLBI experiment will: (i) allow us to measure the size and the expansion of a GRB up to a redshift $z\simeq 0.25$ (at a confidence level of $3\sigma$); (ii) constrain the size $\gtrsim$2 times better than the current global-VLBI array; (iii) improve the localisation precision in Declination from 4 to 30 times; (iv) detect the apparent proper motion of GRBs seen slightly off-axis with a confidence level 3 times better than current VLBI networks. 
Ultimately, the SKA-Mid will open a new window on a portion of the GRB population that has been inaccessible so far.
}

\begin{document}
\maketitle

\section{Introduction}
\label{sec:intro}
Gamma-ray bursts (GRBs) are the most extreme explosions known in the Universe: detected as brief flashes of $\upgamma$-rays (keV--MeV), they last from a fraction of second up to a few hundreds of seconds, during which a significant amount of isotropic-equivalent energy (ranging from $10^{48}$ to $10^{54}$\,erg) is released. Occurring at cosmological distances
, these transients mark the death of an isolated massive star (long-duration GRBs), or the merger of two neutron stars or a neutron star and a black hole in a compact binary system (short-duration GRBs). Regardless of the distinct production channels, both scenarios lead to the formation of a highly magnetised neutron star or a spinning, stellar-mass black hole. This central engine is believed to accrete material and launch two oppositely directed relativistic jets, which then expand into the material surrounding the burst. As the jets expand, the GRB outflow is decelerated by the interstellar medium and an external shock develops, accelerating electrons to relativistic energies. These electrons emit via synchrotron, producing the long-lived afterglow emission that is detected from $\upgamma$-rays to the radio band. 

\begin{figure}[t]
    \centering
	\includegraphics[width=\textwidth]{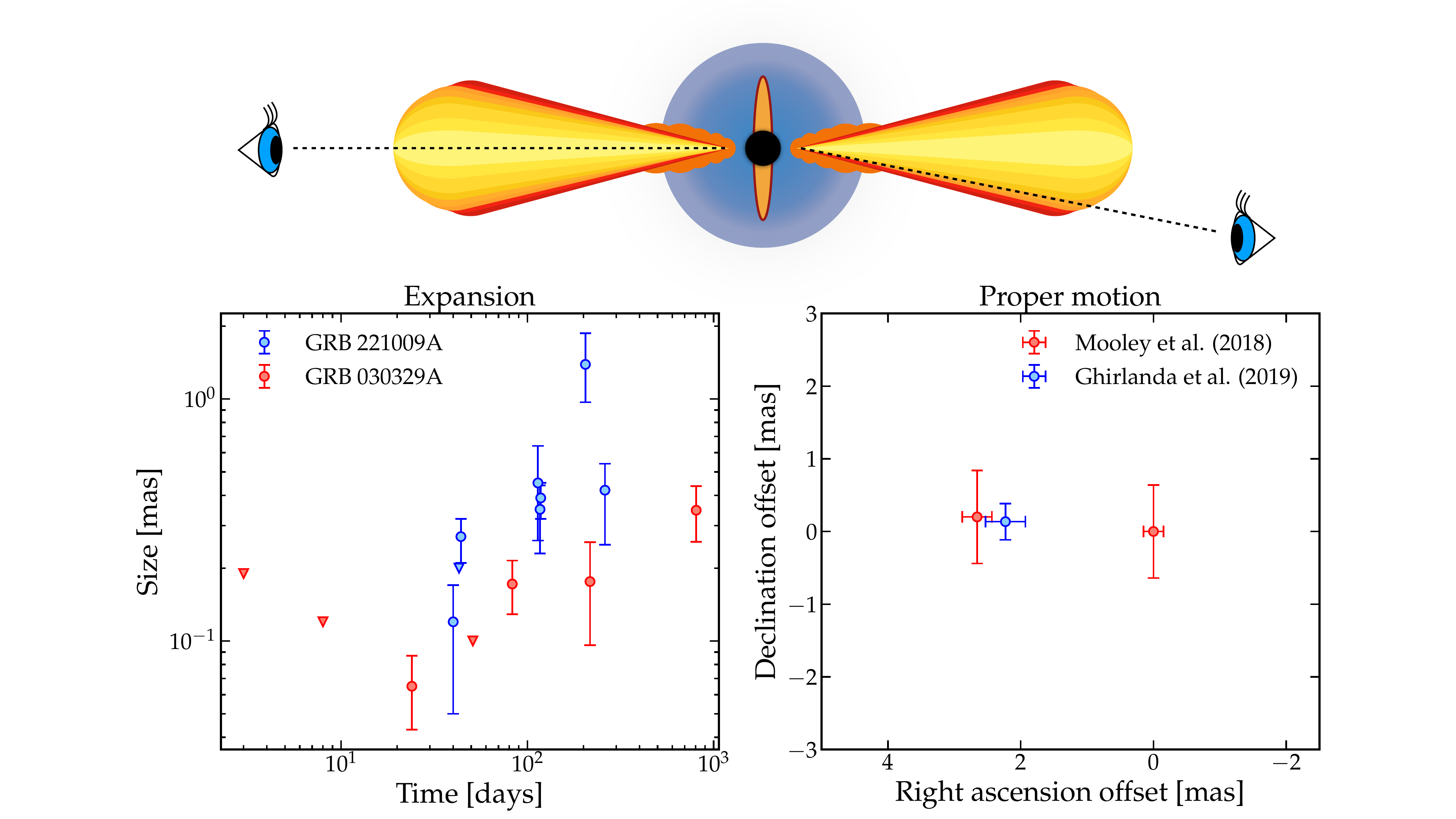}
    \caption{\textit{Left panel}: measurements of the expansion of GRB~030329A \citep{Taylor2004, Taylor2005, Pihlstrom2007} and GRB~221009A \citep{Giarratana2024}. \textit{Right panel}: measurements of the proper motion of GW170817 / GRB~170817A \citep{Mooley2018b, Ghirlanda2019}. A sketch, representative of the two geometrical configurations, is shown on top of the panels.}
    \label{fig:intro}
\end{figure}

The Very Long Baseline Interferometry (VLBI) technique is a unique asset that offers complementary pieces of information on the physics of GRBs that cannot be derived with any other methods or bands to date. Specifically, the high angular resolution ($\sim$milli-arcsecond) provided by VLBI allows us to measure the size and the position of a GRB with great precision. If the GRB jet is pointing (roughly) towards us (on-axis GRB), the size expansion of the blast wave can be measured (left panel of Fig.\,\ref{fig:intro}).
Conversely, if our line of sight is misaligned with respect to the axis of the GRB jet (slightly off-axis GRB), the expansion of the blast wave results into an apparent proper motion of the centroid of the emitting region (right panel of Fig.\,\ref{fig:intro}). 
VLBI provided the first direct evidence of the apparent super-luminal expansion in GRBs \citep{Taylor2004}. For GW~170817 / GRB~170817A, the only event jointly detected both with gravitational and electromagnetic waves to date, the slowly rising flux ($t^{+0.8}$), unveiled by the radio observations between 1 and 7 months, invoked an angular structure of the jet, if present \citep{Mooley2018a}. Alternatively, the same temporal behaviour could be explained by an isotropic outflow with a distribution of its radial velocity and energy profile. The two models could be distinguished by exploiting the high angular resolution of the High Sensitivity Array (HSA; \citealt{Mooley2018b}) and the global-VLBI array \citep{Ghirlanda2019}, which confirmed for the first time ever that the merger of two neutron stars may result into successful jets. 

While these discoveries are of the uttermost importance to characterise the dynamics, the geometry and the structure of the relativistic blast wave, the expansion of a GRB has been measured only for two GRBs hitherto, namely GRB~030329A \citep{Taylor2004, Taylor2005, Pihlstrom2007} and GRB~221009A \citep[left panel of Fig.\,\ref{fig:intro}; ][]{Giarratana2024}, while the apparent motion of the centroid of the emission has been measured only for GRB~170817A / GW170817 \citep[right panel of Fig.\,\ref{fig:intro}; ][]{Mooley2018b, Ghirlanda2019}.
This dearth of useful measurements is mainly due to the combined action of the limited sensitivity of current radio facilities and the paucity of nearby GRBs that are bright enough in radio to be tracked up to very late times. The advent of the SKA, with its superior sensitivity, is expected to make these studies feasible for a larger portion of the GRB population. In this chapter, we quantify the impact of the SKA-Mid, when combined with a VLBI network, on GRB studies.

\section{The impact of the Square Kilometre Array}
\label{sec:simulations}
To estimate the impact of SKA-Mid on VLBI observations of GRB afterglows, we conducted a series of simulations, 
considering five possible VLBI networks: the European VLBI Network (EVN) together with the \textit{enhanced} Multi Element Remotely Linked Interferometer Network (\textit{e}MERLIN); the Very Long Baseline Array (VLBA); a global-VLBI array which includes EVN, VLBA, the Australian Long Baseline Array (LBA), and the phased-up Karl G. Jansky Very Large Array (VLA); the same global-VLBI array in conjunction with the phased-up SKA-Mid in the AA* configuration and with the phased-up SKA-Mid in the final AA4 configuration. In fact, the SKA-Mid will be delivered in stages: the SKA-Mid in the AA* configuration will comprise 144 dishes (maximum baseline of 108\,km), while the AA4 configuration will include 197 dishes (maximum baseline of 159.6\,km)\footnote{\url{https://zenodo.org/records/16951020}}.

To simulate each VLBI observation, we considered a standard frequency setup with central frequency of 5\,GHz (band 5a for the SKA-Mid\footnote{The SKA-Mid 5a band is currently planned to be between 4.6 and 8.5\,GHz}) and a bandwidth of 512\,MHz (recording rate of 4\,Gbps), which are standard values for current VLBI experiments. Since the radio (1--10\,GHz) afterglow of a GRB is usually faint (e.g. flux density $F_\nu\lesssim 1$\,mJy; \citealp[]{Chandra2012}), we assumed a standard phase-referencing experiment of 24\,hr, with a cycle time of 6 minutes: 4 minutes on the target, and 2 minutes on a hypothetical phase calibrator. We produced synthetic visibilites of an observation in an empty field (i.e., pure noise) for each array with the \texttt{observe} task in the \texttt{ehtim} software\footnote{\url{https://github.com/achael/eht-imaging/tree/main}} \citep{Chael2018}. Baseline-dependent thermal noise was added with the \texttt{add\_th\_noise} parameter\footnote{We note that interstellar scintillation may significantly affect the observed size of the source (see, e.g., \citealt{Rickett1990, Goodman1997, Walker1998, Alexander2019}). However, making an accurate prediction of its effect is extremely challenging, given the intrinsically high variability of this phenomenon. We therefore restrict our analysis to the simplest scenario.}. Information on each antenna, namely the geocentric coordinates of the station and the system equivalent flux densities (SEFDs, in Jy) were taken from the EVN Observation Planner\footnote{\url{https://github.com/bmarcote/vlbi_calculator/blob/7a3b8c5cbb8fc0ab2d1598a5b7a85200794e5df4/src/vlbiplanobs/data/stations_catalog.inp}}. The phase centre was set to Dec = $8.8^{\circ}$, in an attempt to optimise the observability from both north and south hemisphere. We imposed that each antenna be able to observe the source only when it is at an elevation between 20$^{\circ}$ and 80$^{\circ}$, as seen from the telescope site. The resulting $(u,v)$--plane for our simulated observations of pure noise with the global-VLBI (left panel) and the global-VLBI + SKA--AA4 (right panel) is shown in Fig.\,\ref{fig:uv-plane_sigma}. The colour map refers to the sensitivity $\sigma$ of each baseline, defined as:
\begin{figure}[t]
    \centering
	\includegraphics[width=0.9\textwidth]{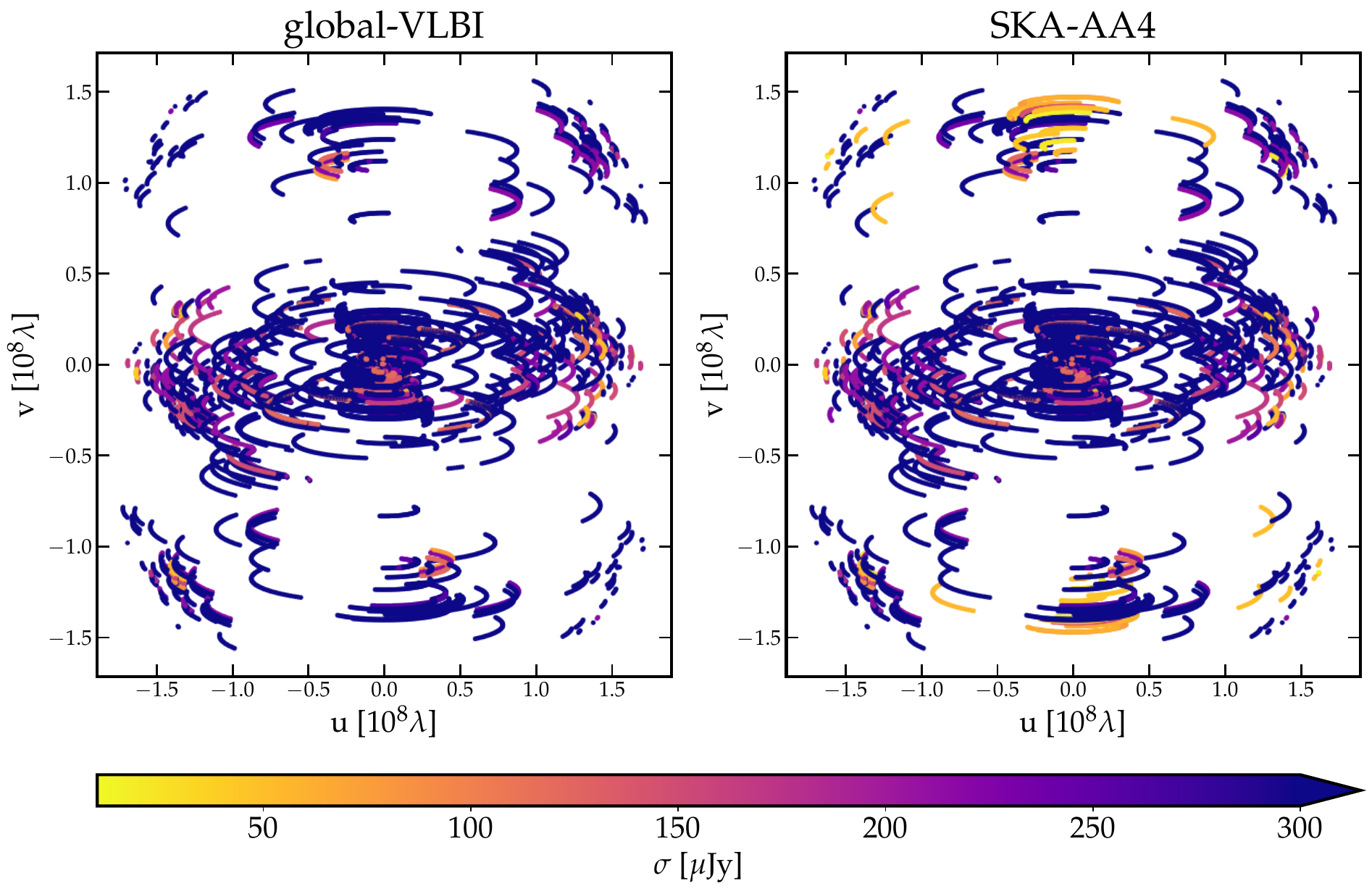}
    \caption{$(u,v)$-plane of our simulated observation with the global-VLBI (left panel) and the global-VLBI + SKA--AA4 (right panel) at 5\,GHz. The colour bar refers to the sensitivity $\sigma$ of each baseline.}
    \label{fig:uv-plane_sigma}
\end{figure}

\begin{equation}
\label{eq:sigma_sefd}
    \sigma = \sqrt{\frac{\mathrm{SEFD_1 \times SEFD_2}}{2\Delta t\Delta\nu}}
\end{equation}

where $\Delta t$ and $\Delta\nu$ are the total time on source and the bandwidth, respectively, while $\mathrm{SEFD}_i$ is the $\mathrm{SEFD}$ of the $i$-th antenna. Fig.\,\ref{fig:uv-plane_sigma} clearly shows that the SKA-Mid in its AA4 configuration will provide baselines that are approximately one order of magnitude more sensitive than the majority of those currently available within the global-VLBI network. The improvement will be particularly significant in the north-south direction.

We represented the GRB afterglow by injecting a model source with a circular Gaussian profile at a random position in the field (RA and Dec close to the phase centre) into the simulated noise visibilities, with a total flux density $F_\nu$ and a given FWHM (mas) of the Gaussian which defines the size of the source. Finally, we fitted a circular Gaussian model to the resulting synthetic visibilities following the Bayesian approach presented in \cite{Salafia2022}. The result is a series of posterior samples of $F_\nu$, FWHM, RA and Dec for each injected source and each array. In the following sections we use the simulation results to demonstrate the impact of the SKA-Mid. 

\subsection{Size measurements}
\label{subsec:expansion}
In this section we focus on the measurement of the size of a GRB. 
First, we selected a bright, on-axis event, and investigated how the precision of the size measurement would change if that GRB were hypothetically placed at higher distances. Second, we explored the $\mathrm{FWHM}$-$F_{\nu}$ parameter space to assess the ability of each VLBI network to measure the size of a GRB with a given flux density and angular size.

\subsubsection{Case study: GRB 221009A}

\begin{figure}[t]
  \centering
  \includegraphics[width=0.49\textwidth]{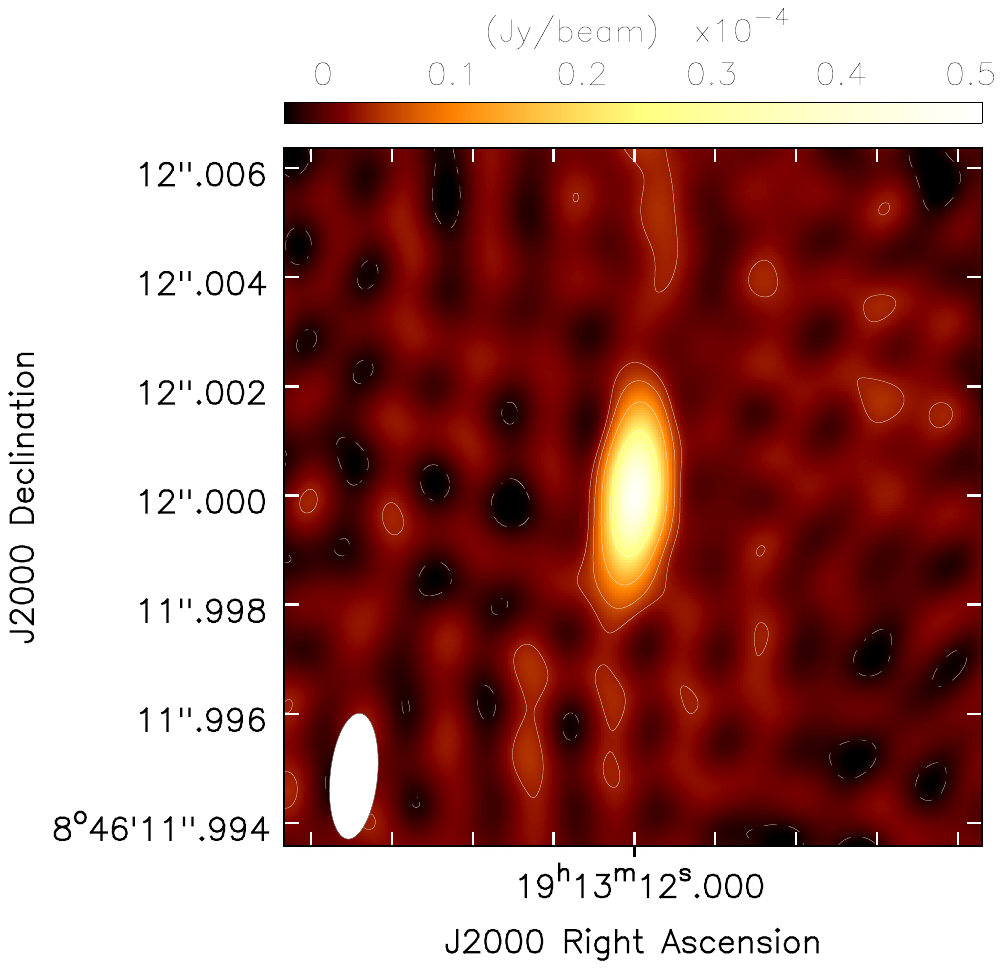}
  \includegraphics[width=0.49\textwidth]{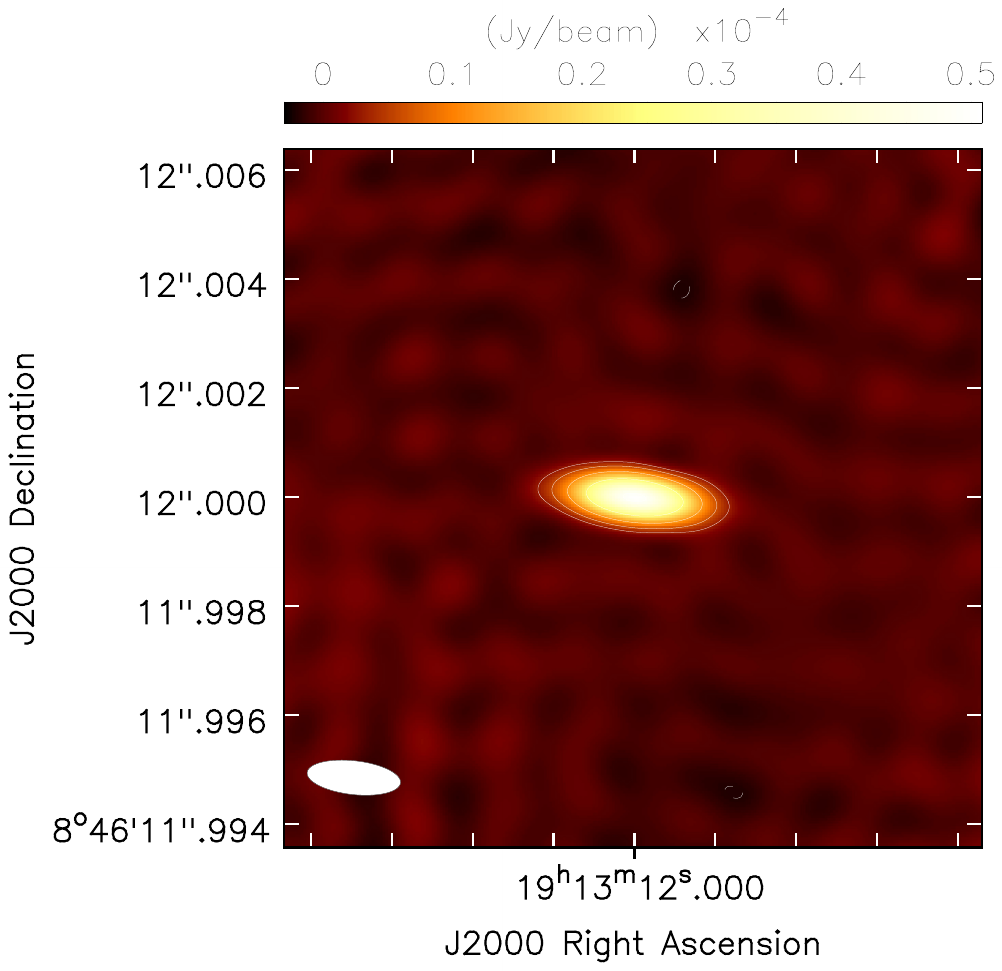}
  \vspace*{-3cm}
  \caption{Simulated images of GRB~221009A as seen with the global-VLBI array (left panel) and the global-VLBI + SKA-Mid in its AA4 configuration (right panel). Contour levels are shown at $-2$ (dashed lines), $3$, $6$, $12$, $24$ and $48\sigma$, where $\sigma=1\,\upmu$Jy/b. The corresponding synthesised beam is shown on the lower left of each image.}
  \label{fig:images}
\end{figure}

We consider the case of GRB~221009A, the brightest GRB ever detected to date. The parameters of the burst are $z_0=0.151$, $F_\nu(z_0)=50\,\mathrm{\upmu Jy}$, $\mathrm{FWHM(z_0)}=0.35\,\mathrm{mas}$ at 100 days post-burst \citep{Giarratana2024}. Fig.\,\ref{fig:images} shows the simulated images for GRB~221009A as seen by the global-VLBI (left panel) and the global-VLBI + SKA-Mid in the AA4 configuration (right panel), respectively. The images were produced with the \textsc{tclean} task in \texttt{CASA} (Version 6.5.4., \citealt{McMullin2007}) using a Briggs weighting scheme with robust parameter set to $0.5$. The inclusion of the SKA-Mid in the VLBI array will clearly improve the resolution in the north-south direction, together with the overall sensitivity of the array. Nevertheless, a source like GRB~221009A will still look unresolved in the image plane, and hence super-resolution will be still needed. The goal of the procedure is to estimate the capability of a given array to measure the size $\mathrm{FWHM}(z)$ of a GRB~221009A-like event located at a given redshift $z$ in the visibility domain. To do so, we simulated the visibilities of such burst as seen at different distances by distinct arrays, as previously explained. The flux density $F_{\nu}(z)$ of the GRB is derived as\footnote{for the \textit{k}-correction we assumed a flat radio spectrum at 100 days post-burst (see, e.g., \citealt{Rhodes2024})}: 

\begin{equation}
\label{eq:flux_density}
    F_\nu(z) = \frac{1+z}{1+z_0}\frac{d_\mathrm{L,0}^2}{d_\mathrm{L}^2}F_\nu(z_0)
\end{equation}

where $d_\mathrm{L}$ is the luminosity distance and $d_\mathrm{L,0} = d_{\mathrm{L}}(z_0)$. The observed size $\mathrm{FWHM}(z)$ is:

\begin{equation}
\label{eq:size}
    \mathrm{FWHM}(z) = \frac{d_\mathrm{A,0}}{d_\mathrm{A}}\mathrm{FWHM}(z_0)
\end{equation}

where $d_{\mathrm{A}}$ is the angular diameter at a given redshift, and $d_{\mathrm{A,0}}=d_{\mathrm{A}}(z_0)$.
We then fitted the synthetic visibilities to retrieve the posterior distribution of the size $\mathrm{FWHM}(z)$.

\begin{SCfigure}[][t]
  \centering
  \includegraphics[width=0.6\textwidth]{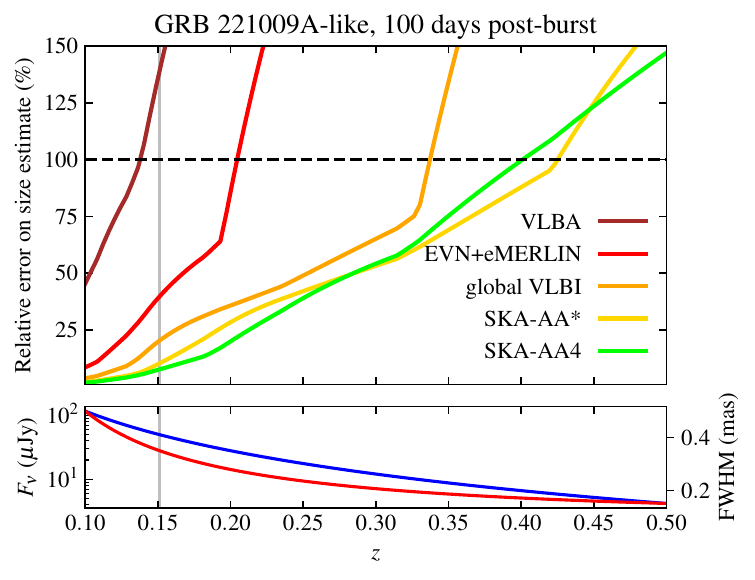}
  \caption{\textit{Top panel:} relative error on the size estimate as a function of distance for a GRB~221009A-like event. Each solid line represents a distinct VLBI network. \textit{Bottom panel:} flux density (blue line, values shown on left-hand y-axis) and FWHM (red line, values shown on right-hand y-axis) of the model source as a function of redshift. In all panels, the grey vertical line shows the real distance of GRB~221009A ($z=0.151$).\\[10pt]}
  \label{fig:size_vs_z}
\end{SCfigure}

To verify whether an array can actually measure the size of the GRB, we defined the relative error on the size estimate as the ratio between the uncertainty $\sigma_{\mathrm{FWHM}}$ (1$\sigma$ confidence level) and the median of the posterior distribution of the size.
Fig.\,\ref{fig:size_vs_z} shows the final results, where each colour refers to a distinct VLBI network. The improvement in the accuracy on the measurement of the size is evident from standard VLBI networks, such as the VLBA and the EVN+{e}MERLIN, to a global-VLBI array. Including the SKA-Mid (both in its AA* and AA4 configurations) in a global-VLBI array provides a significant improvement from $z\simeq 0.32$ on. 
Considering a nominal 3$\sigma$ (5$\sigma$) threshold for a significant measurement of the size, which is equivalent to approximately 33\% (20\%) of relative error, we conclude that a global-VLBI array with the SKA-Mid either in the AA* or AA4 configurations will be able to measure the size of GRB~221009A-like events with a $3\sigma$ ($5\sigma$) confidence level up to a redshift of approximately $z \simeq 0.25$ (0.20) at 100 days post-burst. Moreover, the contribution of the SKA becomes even more evident for relative errors larger than 100\%. At a given redshift, the upper limit on the apparent size provided by a global-VLBI array with the SKA-Mid is expected to be a factor $\gtrsim$2 more constraining that the one provided by a global array without it. Finally, we note that by fitting the VLBI data at each frequency independently,  \citet{Giarratana2024} found different expansion rates, suggesting that distinct components, namely a forward shock and a reverse shock, dominate the emission at different times (but see e.g. \citealt{Rhodes2024}; \citealt{Geng2025}, who draw a different conclusion from long-term, high-cadence monitoring of the GRB light curves and spectra). While this interpretation could not be confirmed owing to the limited signal-to-noise ratio of the VLBI detections, the inclusion of the SKA-Mid in a global-VLBI array has the potential to disentangle the contribution of each component, provided that observations at different frequencies are performed.

\subsubsection{General study}
\begin{figure}[t]
    \centering
	\includegraphics[width=\textwidth]{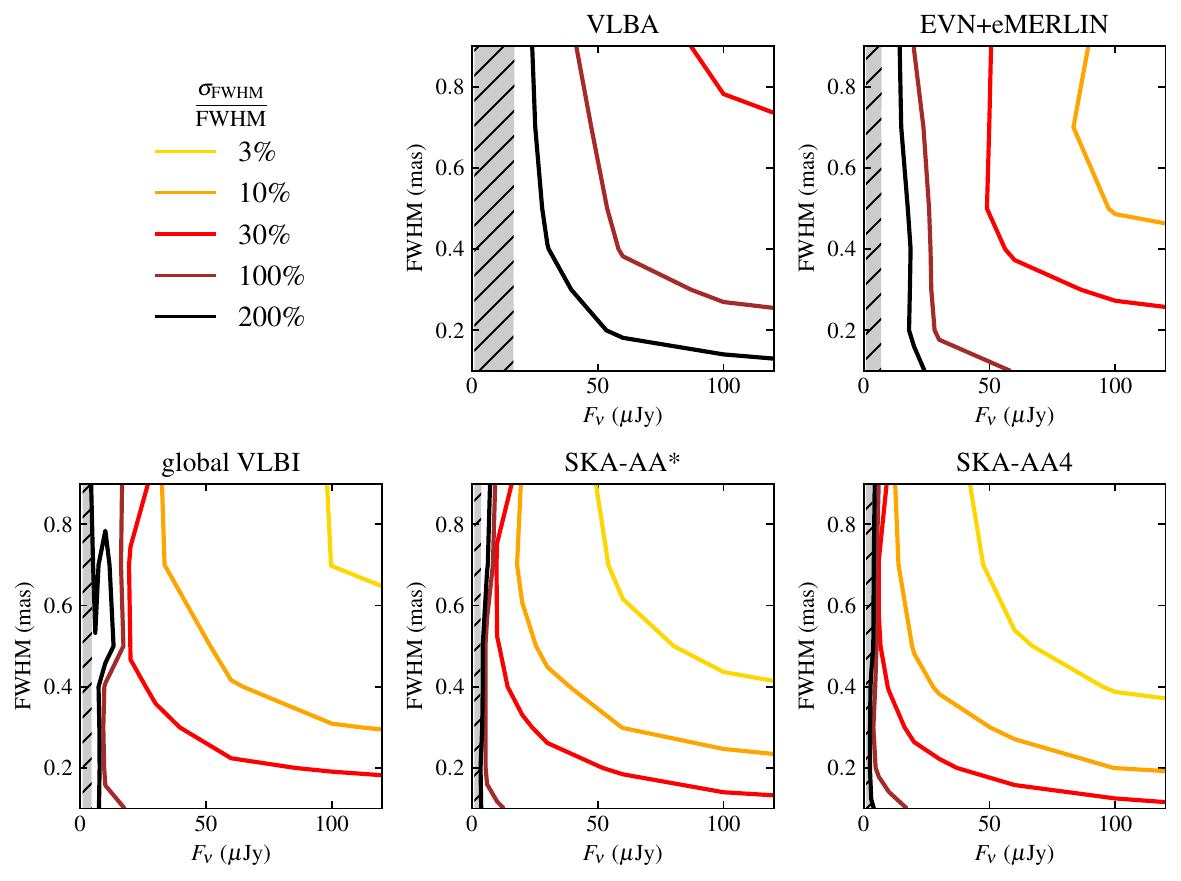}
    \caption{Contour plots of the relative error on the size measurement in the $\mathrm{FWHM}$--$F_{\nu}$ plane. Each panel refers to a distinct array. Solid lines show the contours for a relative error of 3\% (yellow), 10\% (orange), 30\% (red), 100\% (brown) and 200\% (black). The hatched grey area marks the region where the source is not detected because the total flux density is below the $3\sigma_{\mathrm{rms}}$ confidence level, with $\sigma_{\mathrm{rms}}$ being the r.m.s. noise level.}
    \label{fig:resolving_powers}
\end{figure}

Alongside the selected case study, it is important to estimate the capability of a VLBI network to measure the size of a burst with a given size and flux density. To this end, we explored the $\mathrm{FWHM}$-$F_{\nu}$ parameter space by simulating a series of faint, on-axis GRBs. Fig.\,\ref{fig:resolving_powers} shows the contour levels where the combination of the FWHM and $F_{\nu}$ provide a size measurement with a relative error of 3\% (yellow), 10\% (orange), 30\% (red), 100\% (brown) and 200\% (black). 
A point in the bi-dimensional parameter space represents a GRB with given size and flux density; the accuracy on the size estimate of that particular GRB is given by the contour level to which it belongs. For instance, the VLBA can measure the size of a GRB with a relative error less than 30\% only for GRBs with $F_{\nu} \gtrsim$100\,$\upmu$Jy and $\mathrm{FWHM} \gtrsim 0.8$\,mas. On the other hand, EVN + \textit{e}MERLIN can measure the size with similar accuracy for GRBs $F_{\nu} \gtrsim$60\,$\upmu$Jy and $\mathrm{FWHM} \gtrsim 0.4$\,mas. The enhanced $(u,v)$--coverage obtained with a global-VLBI array guarantees an improvement on the relative error of a factor of 3, approximately, for a GRB with the same size and flux density. Finally, a global-VLBI observation including the phased-up SKA-Mid in the AA4 configuration will provide a nominal relative error of 30\% for GRBs with $F_{\nu} \gtrsim$40\,$\upmu$Jy and $\mathrm{FWHM} \gtrsim 0.2$\,mas. 
It is therefore clear that, with the inclusion of the SKA-Mid in the global-VLBI array, we will access a portion of the parameter space that has been inaccessible hitherto. 

To assess the number of events for which it will be possible to measure the size with a $3\sigma$ confidence level, we selected all GRBs detected between 2005 and 2022 with a measured redshift, an estimate of the peak energy $E_p$ of the prompt emission spectrum\footnote{Knowledge of the redshift and $E_p$ allows the energetics of the burst to be estimated.} and at least one radio detection in the 4--8\,GHz band. This selection resulted in a sample of 29 GRBs. For each GRB in the sample, we calculated the projected size following \citet{Granot1999}, assuming a uniform distribution in logarithmic space for the two parameters appearing in their formula $35$: the isotropic-equivalent kinetic energy (assumed to lie between $10^{48}$ and $10^{56}$\,erg) and the circum-burst density (allowed to vary between $10^{-4}$ and $10\,\mathrm{cm}^{-3}$). We then estimated the maximum probability that a given VLBI array can measure the source size with a 3$\sigma$ confidence level by comparing the projected size with the $30\%$ contour levels shown in Fig.\,\ref{fig:resolving_powers}. We find that including the SKA-Mid in its AA4 configuration in a VLBI array would increase the number of resolvable GRBs from approximately 2 to 3 events every $\sim$20 years.

\subsection{Apparent proper motion}
\label{subsec:proper_motion}
Detecting the apparent proper motion provides crucial and independent information on the jet opening angle and the viewing angle, parameters that are extremely difficult to constrain otherwise. In this section, we investigate the capability of a VLBI network to perform these measurements for slightly off-axis GRBs.

\subsubsection{Case study: GRB 170817A / GW170817}
\begin{SCfigure}[][t]
  \centering
  \includegraphics[width=0.6\textwidth]{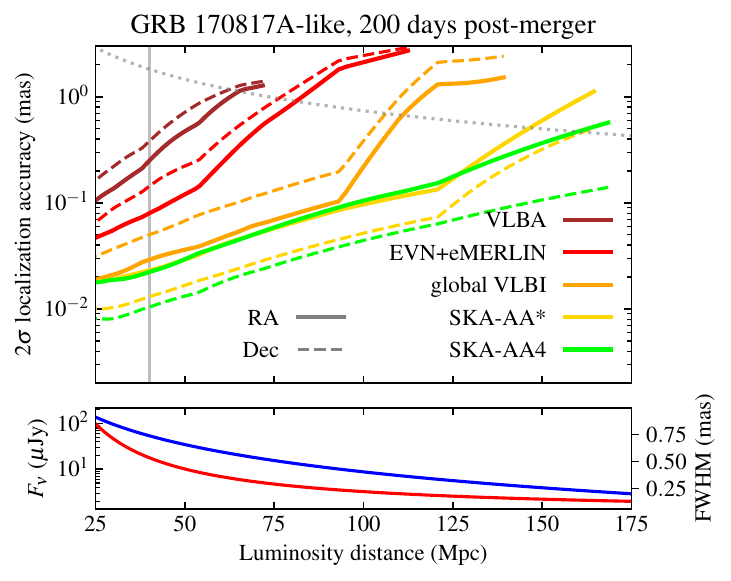}
  \caption{\textit{Top panel:} Localisation accuracy as a function of distance for a GRB~170817A-like event. Each colour represents a distinct VLBI network. The localisation accuracy in right ascension and declination is shown with solid and dashed lines, respectively. The vertical grey line shows the real distance of GRB~170817A (40\,Mpc). The grey dotted line shows the accuracy needed in order to detect a proper motion like the one measured for GRB~170817A, namely 0.017\,mas/d if $z=0.0096$. \textit{Bottom panel}: flux density and size of the model source, same as in Fig.\,\ref{fig:size_vs_z}.}
  \label{fig:pos_vs_z}
\end{SCfigure}

We started from the the famous GW170817 / GRB~170817A. The parameters of the burst are $F_\nu=47\,\mathrm{\upmu Jy}$ and $\mathrm{FWHM}=0.5\,\mathrm{mas}$ at 200 days after the merger \citep{Ghirlanda2019}. The original distance was $d_\mathrm{L,0}=40\,\mathrm{Mpc}$, at a redshift $z_0=0.0096$. The observed proper motion was $\sim 2.7\,\mathrm{mas}$ over $155$ days \citep[between 75 and 230 days post-merger; ][]{Mooley2018b}, which corresponds to an average speed of $0.017$ mas/day or $1.7$ mas in 100 days. Our goal is to estimate the capability of a VLBI network to measure the apparent proper motion of a GRB~170817A-like event placed at different redshifts. The proper motion $\dot\mu (z)$ at a given redshift $z$ scales as:

\begin{equation}
\label{eq:proper_motion}
    \dot\mu(z) = \frac{1+z_0}{1+z}\frac{d_\mathrm{A,0}}{d_\mathrm{A}}\dot\mu(z_0)
\end{equation}

where $\dot\mu (z_0)$ is the apparent proper motion for $z=z_0$. 
We generated synthetic visibilities for a set of GRB~170817A-like events with different redshifts by re-scaling the proper motion (eq.\,\ref{eq:proper_motion}) and the flux density (eq.\,\ref{eq:flux_density}). We fitted the simulated observations with the Bayesian approach developed by \cite{Salafia2022} to retrieve the posterior distributions of the flux density and the position (RA and Dec) of the generated source. Fig.\,\ref{fig:pos_vs_z} presents the $2\sigma$ astrometric precision as a function of distance of the GRB. The coloured lines refer to distinct arrays. Solid and dashed lines show the accuracy in RA and Dec, respectively. The vertical grey line indicates the real distance of GRB~170817A (40\,Mpc), while the grey dotted line is the accuracy needed in order to have a proper motion like the one measured for GRB~170817A, namely 0.017\,mas/d if $z=0.0096$. For instance, if a coloured line, representative of a given array, lies below the grey dotted line at a given distance, that VLBI network will be able to detect an apparent proper motion such as the one observed for GRB~170817A with an accuracy larger than $2\sigma$ at that distance. 

Some important conclusions can be drawn from Fig.\,\ref{fig:pos_vs_z}. First, the localisation accuracy naturally improves for more sensitive arrays, as expected. Furthermore, current VLBI networks provide more sensitive baselines in the E-W direction, resulting in a better accuracy in RA (solid lines). However, when the SKA-Mid joins the experiment, the situation is inverted: as the phased-up SKA-Mid represents the most sensitive antenna in the VLBI network, the north--south baselines become the most sensitive ones. This in turn translates into a greater accuracy in Dec (dashed lines). For this reason, when the SKA-Mid is included in the array the localisation accuracy in RA improves significantly only for distances larger than approximately 90\,Mpc (when the sensitivity plays a major role), while the improvement in Dec is striking at any distances, from approximately 4 to even 30 times better with the SKA-Mid.

\subsubsection{General study}
\begin{figure}[t]
    \centering
	\includegraphics[width=1\textwidth]{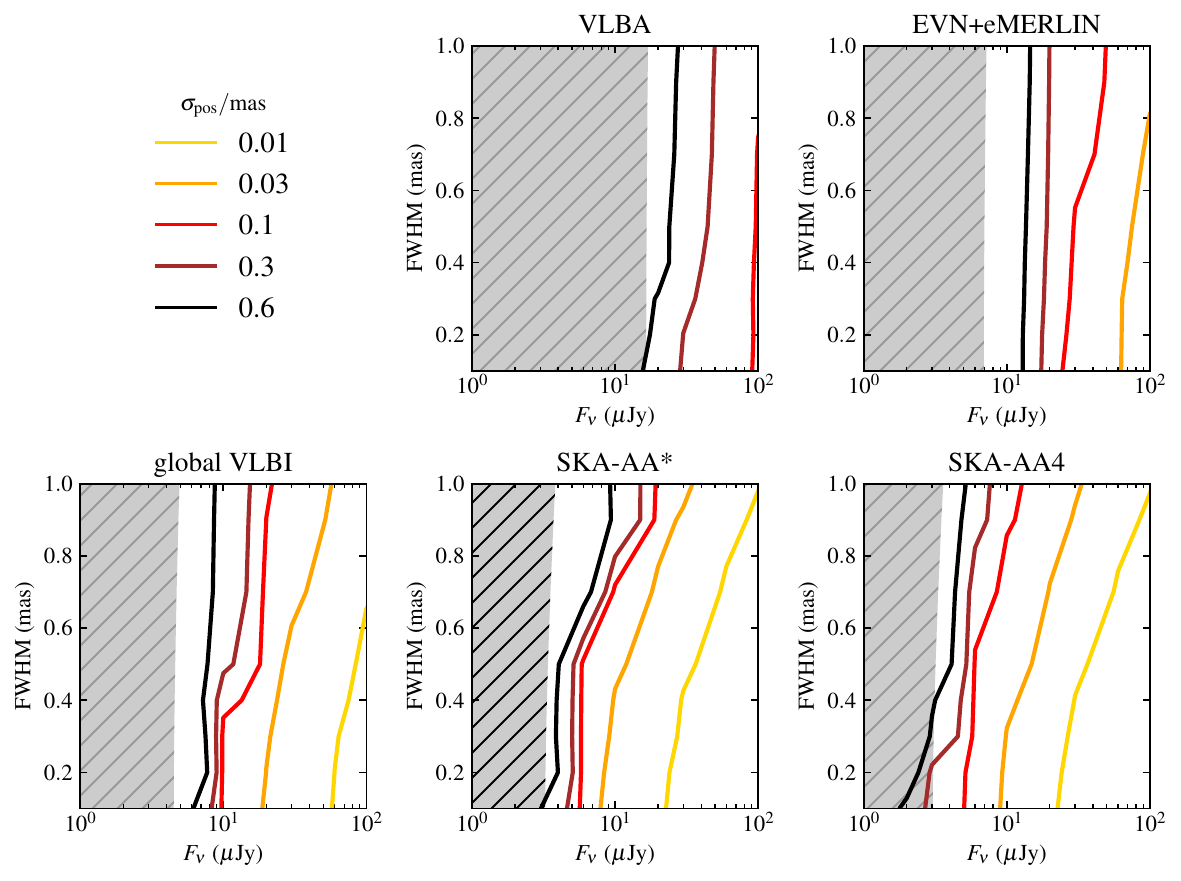}
    \caption{Contour plots of the uncertainty on the position in the $\mathrm{FWHM}$--$F_{\nu}$ plane. Each panel refers to a distinct array. Solid lines show the contours for an error of 0.01 (yellow), 0.03 (orange), 0.1 (red), 0.3 (brown) and 0.6\,mas (black). The hatched grey area marks the region where the total flux density of the source is below the $3\sigma_{\mathrm{rms}}$ confidence level, with $\sigma_{\mathrm{rms}}$ being the r.m.s. noise level.}
    \label{fig:pos_powers}
\end{figure}
In order to estimate the nominal astrometry accuracy each array can achieve, we performed a series of simulations following the procedure already presented. The final aim is to assess whether the nominal localisation accuracy of the VLBI networks is sufficient to measure the apparent proper motion of a slightly off-axis GRB with given flux density and size\footnote{While novel techniques, such as the MultiView approach \citep[e.g.,][]{Rioja2017}, can significantly improve the astrometry, phase de-correlation in standard phase-referencing experiment can considerably deteriorate the localisation accuracy. Therefore, we opted for a simplified approach, limiting our analysis to the nominal localisation uncertainty of each array.}. We defined the localisation accuracy as

\begin{equation}
\label{eq:localisation}
    \sigma_{\mathrm{pos}} = 2 \sqrt{\sigma_{\mathrm{RA}}^2 + \sigma_{\mathrm{Dec}}^2}
\end{equation}

where $\sigma_{\mathrm{RA}}$ and $\sigma_{\mathrm{Dec}}$ are the standard deviations of the posterior distributions of the position in RA and Dec, respectively. Fig.\,\ref{fig:pos_powers} shows the contour level of $\sigma_{\mathrm{pos}}$ in the $\mathrm{FWHM}$--$F_{\nu}$ plane. 
Solid lines show the contours for an error of 0.01 (yellow), 0.03 (orange), 0.1 (red), 0.3 (brown) and 0.6\,mas (black). The hatched grey area marks the region of the bi-dimensional space where the source is not detected because its total flux density is below the $3\sigma_{\mathrm{rms}}$ confidence level, with $\sigma_{\mathrm{rms}}$ being the r.m.s. noise level. 

A minimum of two measurements of the position of the source in time are required to estimate the proper motion. In the bi-dimensional space, this results into two points: the same GRB, but different size and flux density, as these quantities evolve with time. For each of the two epochs, the flux density and the size of the GRB will define a localisation accuracy for a given array. Assuming that there are two epochs with same localisation accuracy $\sigma_{\mathrm{pos}}$ (namely, they lie on the same contour level on the $\mathrm{FWHM}$--$F_{\nu}$ plane), a proper motion is detected with a 1.4$\sigma$ confidence level if the displacement observed between the two epochs is 2$\sigma_{\mathrm{pos}}$. If the observed displacement is larger, the confidence level for the proper motion will also increase.

It is evident from Fig.\,\ref{fig:pos_powers} the astonishing impact that the SKA will have. For instance, let us assume a compact (i.e., unresolved) source with flux density of 20\,$\upmu$Jy. An observation with a global-VLBI array gives an expected localisation accuracy of $\sigma_{\mathrm{pos}}=0.03$\,mas. Taking the observed proper motion of GRB~170817A as a reference, after 20 days the displacement would be of 0.4\,mas. This displacement would be detected with a 9$\sigma$ confidence level with the global-VLBI array. Conversely, if the SKA-Mid joins the experiment, the confidence level increases by approximately a factor of 3.

\section{Beyond the Square Kilometre Array}
\label{sec:improvements}
The SKA will significantly enhance our ability to investigate the dynamics, geometry, and structure of the GRB relativistic blast waves with unprecedented precision. Although our simplified simulations have demonstrated the remarkable impact of the SKA when included in a VLBI network, we note that there is a growing global effort to further optimise both the frequency setups and the network of antennas that will participate in future VLBI experiments. Here we explore how near-term upgrades would multiply the science return from GRB VLBI imaging, including higher-frequency coverage, denser global baselines, advanced digital backends, and strategic network expansions.

\subsection{Higher Frequencies: global-VLBI with Band 5b}

In our current analysis, we have considered only the SKA Band 5a (4.6--8.5~GHz) within a standard VLBI observation. However, SKA Band 5b (8.3--15.4~GHz) is also planned to be available and offers two immediate advantages for GRB afterglow studies. First, the higher frequency provides \textbf{improved angular resolution}, approximately 2--3$\times$ better than Band 5a, allowing us to detect centroid motions earlier in the afterglow's evolution. Second, the broader bandwidth available in Band 5b results in \textbf{enhanced sensitivity}, since the r.m.s. noise level scales as $\propto \Delta\nu^{-1/2}$. This is especially useful at late times, when the afterglow emission has faded. 

At present, the EVN and the LBA can observe at 8.3~GHz (X-band) with limited bandwidth, while only the VLBA can observe up to 15~GHz with a bandwidth of 3.4~GHz\footnote{\url{https://science.nrao.edu/facilities/vlba/docs/manuals/oss/bands-perf}}. However, promising VLBI experiments are testing the performance of the EVN at these frequencies, and if existing VLBI networks are equipped with receivers capable of covering the SKA Band 5b in the near future, this broader frequency range will enable deeper, faster detections. Moreover, while the synchrotron emission from a GRB afterglow fades and progressively shifts towards lower frequencies over time, the central frequency of Band 5b remains low enough to ensure that the afterglow will be detectable long enough to track size evolution and/or apparent proper motion of the emission centroid. In this sense, Band 5b represents an optimal frequency window for these studies. 

\subsection{Filling the $(u,v)$-plane: Africa, East Asia, and Oceania}

The reconstruction of high-fidelity VLBI images is heavily dependent on $(u,v)$-plane coverage. The chronic lack of intermediate north--south baselines ($\sim$10$^{3}$--10$^{4}$~km) in current VLBI networks reduces final image fidelity and worsens the overall achievable r.m.s. noise level. 
The \textbf{African VLBI Network} (AVN; \citealt{Bempong-Manful01.2026.SKA}) represents a natural solution to fill the gap between inter-European baselines on the one hand and the longer baselines with South African and Australian antennas on the other. The presence of a very sensitive element such as the phased-up SKA in the AA4 configuration will also guarantee better calibration for these smaller antennas in Africa, enhancing sensitivity across the array and improving de-convolution stability while reducing beam ellipticity.

A further strategic step involves the inclusion of a new synthesised array around the Five-hundred-meter Aperture Spherical Telescope (FAST), called \textbf{FAST Core Array} \citep{Jiang2024}. The FAST Core Array, consisting of the 500-m FAST telescope and approximately 64 newly built 40-m antennas, will form the most sensitive centimetre-wavelength interferometer in the northern hemisphere. Operating across a broad frequency range of 0.3–10\,GHz, the 40-m antennas will offer powerful synergy with the SKA-Mid and the global-VLBI array. Specifically, high signal-to-noise ratio baselines from Chinese stations would be particularly important for improving calibration of existing East Asian VLBI elements and, crucially, for the LBA, whose calibration quality is often marginal for the faintest GRB afterglows. Furthermore, long, sensitive baselines between China and the LBA would improve the $(u,v)$--plane coverage, substantially boosting the dynamic range by reducing sidelobe artifacts in the final image. 

\subsection{Advances in Digital Backends and Rapid Response}
Several system-level improvements would immediately benefit VLBI imaging of GRBs. First,
transitioning from 4~Gbps (512~MHz bandwidth) to 16--32~Gbps (2--4~GHz bandwidth) recording rates would lower thermal noise by a factor of $\approx$2--3, allowing us to track the afterglow evolution up to later times. Moreover, dual-band, co-timed observations in Bands 5a and 5b with multi-frequency phase transfer would improve coherence on the longest baselines by stabilizing high-frequency phases, which has been proven crucial for imaging compact GRB afterglows with angular sizes $\lesssim$1~mas. Furthermore, using multiple phase calibrators reduces residual tropospheric and ionospheric systematics, ensuring relative astrometry precision at the $\lesssim$10--30~$\upmu$as level necessary for detecting proper motion in slightly off-axis jets. Finally, routine fibre-connected e-VLBI with sub-array triggers would allow observations within 12--24 hours post-burst, capturing key reverse-shock peaks and early superluminal motion that are otherwise missed.


\subsection{A truly global-VLBI network for GRBs}
By integrating Band 5b operations, AVN deployment, a sensitive Chinese array anchored by FAST, high-rate backends with multi-band phase transfer, and routine e-VLBI, we envision establishing a global-VLBI facility to investigate GRBs, with the SKA being its southern anchor. This network is expected to (i) deliver systematic early-epoch imaging for a larger fraction of well-localised GRBs; (ii) provide robust size and proper motion measurements with $\upmu$as-level systematic uncertainties; (iii) push resolved imaging and proper-motion measurements to higher redshifts. These improvements will, in turn, provide tighter constraints on the magnetisation of the GRB ejecta, enable population-level tests of the GRB structure (structured versus chocked jet models), allow us to study jet physics across cosmic time.

\section{Conclusions}
\label{sec:conclusions}
VLBI observations of GRBs are crucial to draw direct information that cannot be retrieved with any other methods or bands to date. Specifically, measuring the size and the expansion of a GRB allows us to characterise the dynamics (apparent superluminal velocity) and structure (structured vs chocked jet) of the blast wave. On the other hand, the detection of apparent proper motion in slightly off-axis GRBs provides direct constraints on the viewing angle and the jet opening angle, alleviating or even breaking the degeneracy in the modelling of GRB afterglows.
In this chapter, we quantified the impact that the SKA-Mid will have on VLBI studies of GRBs. Involving the SKA-Mid in a global array will, in fact, improve both the resolution and the localisation precision in an unprecedented way.

In order to estimate this improvement, we employed a set of dedicated simulations of VLBI observations of GRBs. We considered five different VLBI networks. We showed that including the SKA-Mid in a global-VLBI observation will extend the current limit for which we can measure the size of a GRB to $z \simeq 0.25$ (at a confidence level of $3\sigma$). For GRBs located at larger distances, a global-VLBI network with the SKA-Mid is anticipated to provide upper limits on the size that are a factor $\gtrsim$2 more constraining that the ones provided by a global-VLBI array without it. Therefore, the inclusion of a very sensitive antenna such as the phased-up SKA-Mid will open a new window on a portion of the GRB population that has been inaccessible so far.
The SKA-Mid will also have a tremendous impact on the localisation accuracy. Since the SKA-Mid will compensate for the lack of long, sensitive baselines in the north--south direction, including this facility in a global-VLBI experiment will improve the astrometry in Dec by a factor of 4 to 30. This translates into a striking capability to detect the apparent proper motion of slightly off-axis GRBs: with the SKA-Mid, the confidence level for these studies is expected to increase by a factor of 3 with respect to measurements performed by current VLBI networks.

To conclude, the SKA-Mid will represent the southern hemisphere anchor of a truly global-VLBI facility, which will allow us to routinely investigate these catastrophic events with unmatched sensitivity, resolution and astrometry accuracy, charting the future of GRB studies and establishing VLBI as an indispensable tool for understanding the most extreme explosions in the Universe.

\bibliographystyle{abbrvnat-maxbibnames4}
\bibliography{chapter} 

\end{document}